\documentclass[11pt]{article}
\usepackage{amsfonts,epsfig,latexsym}

\newcommand{\R}{{\mathbb{R}}}
\newcommand{\Z}{{\mathbb{Z}}}

\newcommand{\beq}{\begin{equation}}
\newcommand{\eeq}{\end{equation}}
\newcommand{\bea}{\begin{eqnarray}}
\newcommand{\eea}{\end{eqnarray}}
\newcommand{\ra}{\rightarrow}

\newcommand{\wt}{\widetilde}

\newcommand{\et}{\widetilde{{\cal E}}}

\begin{document}

\title{A quantum Peierls-Nabarro barrier}
\author{J.M. Speight\thanks{E-mail: {\tt speight@amsta.leeds.ac.uk}}\\
Department of Pure Mathematics, University of Leeds\\
Leeds LS2 9JT, England}

\date{}

\maketitle
\begin{abstract}
Kink dynamics in spatially discrete nonlinear Klein-Gordon systems is 
considered. For special choices of the substrate potential, such systems
support continuous translation orbits of static kinks with no (classical)
Peierls-Nabarro barrier. It is shown that these kinks experience, 
nevertheless, a lattice-periodic confining potential, due to purely quantum
effects anaolgous to the Casimir effect of qunatum field theory. The
resulting ``quantum Peierls-Nabarro potential'' may be calculated in the 
weak coupling approximation by a simple and computationally cheap numerical
algorithm, which is applied, for purposes of illustration, to a certain
two-parameter family of substrates. 
\end{abstract}

\section{Introduction}
\label{int}

Many systems in condensed matter and biophysics may be modelled by infinite
chains of coupled anharmonic oscillators. If the anharmonic
substrate potential has two or more degenerate vacua, such a system may
support
static kink solutions interpolating between neighbouring vacua. These kinks
have various interesting physical interpretations (as crystal dislocations
\cite{disloc}, charge density waves \cite{cdwave} and magnetic \cite{magdom}
and ferroelectric \cite{ferdom} domain walls, for example)  
and their dynamics is an interesting and important subject.

Such a system of oscillators has an alternative interpretation as a
spatially discrete version of an appropriate nonlinear Klein-Gordon equation.
The lattice spacing $h$ is related to the spring constant of the chain
$\alpha$ by $\alpha=1/h^2$ so the strong spring-coupling
 limit is interpreted as a 
continuum limit. In the continuum limit, static kinks may occupy any position
in space, by translation symmetry. This is generically untrue in the discrete
system: static kinks may generically occupy only two positions relative to 
the lattice, one of which is a saddle point of potential energy, the other
a local minimum. The difference in energy between these two static solutions
is the Peierls-Nabarro (PN)
barrier, the barrier  which a kink must surmount in 
order to propagate from one
lattice cell to the next. It can have strong effects on the dynamics of kinks
in the system (kink trapping, radiative deceleration, phonon bursts etc.\
\cite{effects1,effects2}).

One might expect the PN barrier, and hence its effects, to grow monotonically
with $h$ (of course the barrier vanishes as $h\ra 0$). However, recent work
of Flach, Kladko and Zolotaryuk \cite{flach}
has shown that this is certainly not 
universally true. In fact, there exist infinitely many substrate potentials
with the property that at at least one none-zero lattice spacing, $h_*$ say,
the PN barrier vanishes exactly and a continuous translation orbit of static
kinks is recovered. We shall say that a substrate potential with this 
property is ``transparent at lattice spacing $h_*$.'' Such potentials may be
constructed by means of the so-called Inverse Method, and are clearly of some
theoretical interest. 

The purpose of the present paper is to argue that although the kinks of such
a system (at $h=h_*$) are free of the classical PN barrier, they still 
experience a qualitatively similar periodic confining potential due to
quantum effects analogous to the Casimir effect of quantum electrodynamics. 
We call this the quantum Peierls-Nabarro (QPN)
potential. The essential physical
observation is that the total zero-point energy of the lattice phonon modes
in the presence of a kink depends periodically on the kink position. It
should be emphasized that the kink position itself is treated as a classical
degree of freedom while the phonons are quantized. The physical regime in 
which this is consistent will be identified: the classical kink mass must far
exceed the phonon mass, and the kink must interpolate between widely 
separated vacua. For purposes of illustration, we shall compute the QPN
potential numerically for a two parameter family of substrate potentials
which (in a sense) includes discrete sine-Gordon and $\phi^4$ systems. As
a by-product of these calculations, we will obtain numerical evidence in
favour of the assumption that kinks in these models are classically stable.

\section{Construction of transparent substrate potentials by the Inverse
Method}
\label{inverse}

The general discrete nonlinear Klein-Gordon system consists of a field 
$\phi:\Z\times\R\ra\R$ whose evolution is determined by a second order
differential difference equation,
\beq
\label{eleq}
\ddot{\phi}_n=\frac{1}{h^2}(\phi_{n+1}-2\phi_n+\phi_{n-1})-V'(\phi_n).
\eeq
Here $h$ is the spatial lattice spacing, $\ddot{\phi}_n=d^2\phi_n/dt^2$ 
and $V$
is the substrate potential. One interpretation of the equation of motion
is as that of an infinite system of identical oscillators (each oscillating
in potential well $V$) with nearest neighbours coupled by identical Hooke's
law springs of strength $\alpha=h^{-2}$. As the name suggests, (\ref{eleq})
becomes a nonlinear Klein-Gordon equation 
in the continuum limit, $h\ra 0$. 
If $V(\phi)$ has neighbouring degenerate vacua
at $\phi=a_-$ and $\phi=a_+>a_-$, say, (\ref{eleq})
supports static kinks interpolating between them. To find these requires the
solution of a second order nonlinear difference equation subject to the
boundary conditions $\lim_{n\ra\pm\infty}\phi_n=a_\pm$, which is usual only 
possible numerically. 

In this section we will construct, for a given lattice spacing $h_*>0$, a 
substrate potential $V_{h_*}(\phi)$ which supports a continuous translation
orbit of static kinks, and so by construction is transparent at lattice
spacing $h_*$. To do this we shall use a variant of the Inverse Method of
Flach, Kladko and Zolotaryuk \cite{flach,tdk} (which was originally devised
to construct substrate potentials which support exact {\em propagating},
rather than static, kink solutions). The idea is to {\em choose} a static
kink profile, that is an analytic, monotonic surjection $f:\R\ra(a_-,a_+)$,
satisfying exponential decay critera as $z\ra\pm\infty$,
\beq
\label{3.5}
f(z)=a_\pm+O(e^{-\mu|z|}),\qquad \mu>0,
\eeq
and the symmetry requirement,
\beq
\label{3.6}
f(z)+f(-z)\equiv a_++a_-,
\eeq
then {\em impose} that the translated kink $\phi^b_n=f(nh-b)$ be a static
solution of the system (with $h=h_*$) for all $b\in\R$. From
equation (\ref{eleq}), this condition holds
 provided one chooses $V_{h_*}$ such that
\beq
\label{4}
V_{h_*}'(f(z))=\frac{1}{h_*^2}[f(z+h_*)-2f(z)+f(z-h_*)]
\eeq
for all $z\in\R$. This uniquely determines $V_{h_*}':(a_-,a_+)\ra\R$ by
monotonicity of $f$, and hence $V_{h_*}:(a_-,a_+)\ra\R$ up to an arbitrary
constant. To complete the definition of this transparent substrate, one 
should extend its definition appropriately to all $\R$. How one does this is
somewhat arbitrary, but will have no bearing on our results, so we shall
merely demand that $V_{h_*}$, $V_{h_*}'$, $V_{h_*}''$ be continuous at
$a_\pm$. Equations (\ref{3.5}) and (\ref{4}) then imply
\bea
\label{5}
V_{h_*}'(a_\pm)&=&0 \\
\label{6}
V_{h_*}''(a_\pm)&=&\frac{2}{h_*^2}(\cosh\mu h_*-1)>0.
\eea
Hence $\phi=a_\pm$ are both stable equilibria. To see that these are 
degenerate vacua, note that from (\ref{4}),
\bea
h_*^2[V_{h_*}(a_+)-V_{h_*}(a_-)]&=&\int_{-\infty}^\infty[f(z+h_*)-2f(z)+
f(z-h_*)]f'(z)\, dz \nonumber \\
&=&\int_{-\infty}^\infty[f(z+h_*)+f(-(z+h_*))]f'(z)\, dz -(a_+^2-a_-^2)
\nonumber \\
&=&(a_++a_-)[f(z)]^\infty_{-\infty}-(a_+^2-a_-^2) \nonumber \\
&=&0,
\eea
using the symmetry constraint on $f$, (\ref{3.6}).

So given a kink profile $f$, the inverse method generates a one-parameter
family of double-well substrate potentials $\{V_{h_*}:h_*>0\}$ each of which
is transparent at spacing $h=h_*$. Moreover, one has as explicit formula for
the continuous translation orbit of static kink solutions, namely
$\phi^b_n=f(nh_*-b)$, $b\in\R$.  On physical grounds, one
expects the static kinks to be stable to small perturbations, although
strictly speaking this is not assured. One would need to check that the
Hessian of the potential energy functional,
\beq
E_P=\sum_{n\in\Z}\left[
\frac{1}{2h_*^2}(\phi_{n+1}-\phi_n)^2+V_{h_*}(\phi_n)
\right],
\eeq
 about $\phi^b$ has strictly positive spectrum (except for
the zero-mode associated with translation). The quantum calculation 
described in section \ref{qpnp} may be reinterpreted as the calculation of
this spectrum. The results of section \ref{numerics} then
constitute numerical confirmation of kink stability for the family of
transparent substrates considered therein.

\section{The quantum Peierls-Nabarro potential}
\label{qpnp}

In this section we shall quantize the system using a weak coupling 
approximation, essentially following the method outlined in \cite{raj}, 
adapted to the infinite lattice. The method has previously been applied in 
the spatially discrete context to a certain nonstandard lattice sine-Gordon
model \cite{kce}.

One may regard $E_P$ as a potential energy function on the inifinite 
dimensional space $Q$ of sequences $\phi:\Z\ra\R$. The vacua $\phi=a_\pm$ 
lie at the bottom of identical potential wells, cut off from each other, and
all configurations with kink boundary behaviour, by an infinite energy 
barrier. Assuming that $V$ is transparent at the lattice spacing under 
consideration (e.g.\ $V=V_{h_*}$ and $h=h_*$, as in section \ref{inverse}),
the continuous kink translation orbit is an equipotential curve in $Q$: the
classical energy of a static kink is independent of its position. If
the kinks are stable, this is a level valley bottom winding through $Q$.

Quantum mechanically, a particle cannot sit at the bottom of a potential 
well, or on the low dimensional floor of a valley: it always possesses a
zero-point energy dependent on the shape of the well bottom. In this section
we will semi-classically quantize motion both in the vacuum and kink sectors 
of the system. A physical regime will be identified in which the kink is
very heavy, so that the kink position $b$ may be treated as a classical
degree of freedom, while the comparatively light phonon modes orthogonal
to the translation mode are quantized perturbatively, by Taylor expansion
of $E_P$. Computation of the kink
ground state energy then amounts to summing the
zero-point energies of an infinite system of harmonic oscillators, resulting
in a divergent series. The quantity of physical significance is not this
energy, but rather the difference between the kink and vacuum ground state
energies, which is expected to be finite. Since translation is not a 
symmetry of the discrete system, there is no reason to expect this energy
to be independent of $b$, that is, one expects the quantum kink energy
to vary periodically with kink position, which is the origin of the 
quantum PN potential.

It is convenient to introduce a dimensionless coupling constant $\lambda$
into the model so that the (classical) Hamiltonian of the system is
\beq
\label{10}
H=\sum_{n\in\Z}\left[\frac{1}{2}\pi_n^2+\frac{1}{2h^2}(\phi_{n+1}-
\phi_n)^2+\frac{1}{\lambda^2}V(\lambda\phi)\right]
\eeq
where $\pi_n=\dot{\phi}_n$ is the momentum conjugate to $\phi_n$. Assuming
that $V$ is transparent at spacing $h$, this system supports a continuous
translation orbit of static kinks
\beq
\label{11}
\phi^{b,\lambda}_n=\frac{1}{\lambda}f(nh-b),\qquad b\in\R
\eeq
interpolating between $a_-/\lambda$ and $a_+/\lambda$. The classical energy
of these kinks is independent of $b$ and clearly scales with $\lambda$ as
\beq
\label{12}
E_P[\phi^{b,\lambda}]=\frac{1}{\lambda^2}E_P[\phi^{0,1}].
\eeq
The physical regime of interest is that of small $\lambda$ where the kinks
interpolate between widely separated vacua, by (\ref{11}), and are very 
heavy, by (\ref{12}). In this regime, one may approximate motion about any
stable static solution $\wt{\phi}=\varphi/\lambda$
by using a truncated Taylor series approximation for
$E_P[\wt{\phi}+\delta\phi]$:
\beq
\label{13}
E_P[\wt{\phi}+\delta\phi]=\frac{1}{\lambda^2}E_P[\varphi]+
\frac{1}{2h^2}\sum_{n,m}W_{nm}\delta\phi_n\delta\phi_m+O(\lambda)
\eeq
where
\beq
\label{14}
W_{nm}=\delta_{nm}[2+h^2V''(\varphi)]-\delta_{n,m+1}-\delta_{n,m-1}.
\eeq
Since $W$ is a real, symmetric matrix, there exists an orthogonal 
transformation $R$ such that
\beq
\label{15}
W_{nm}=\sum_{k,l}R^T_{nk}U_{kl}R_{lm}
\eeq
with $U$ diagonal. The diagonal entries of $U$ are the eigenvalues 
$\Lambda_n$ of $W$, none of which is negative provided $\wt{\phi}$ is
stable, as we are assuming. Introducing normal coordinates $\xi_n=\sum_m
R_{nm}\delta\phi_m$ which have conjugate momenta $\eta_n=\sum_m R_{nm}\pi_m$,
the Hamiltonian for motion about $\wt{\phi}$ reduces to
\beq
\label{16}
H=\frac{1}{\lambda^2}E_P[\varphi]+\frac{1}{2}\sum_n[\eta_n^2+
\frac{\Lambda_n}{h^2}
\xi_n^2]+O(\lambda).
\eeq
Neglecting the $O(\lambda)$ remainder, this is the Hamiltonian for an 
inifinite set of decoupled harmonic oscillators of natural frequencies
$h^{-1}\sqrt{\Lambda_n}$.

We now quantize in standard canonical fashion in the cases where 
$\wt{\phi}=a_\pm/\lambda$ (the vacuum)
and $\wt{\phi}=\phi^{b,\lambda}$ (the kink
located at $b$). Let the $W$-matrices in these cases be denoted $W^{vac}$
and $W^K(b)$ repectively, with spectra $\{\Lambda^{vac}_n\}$ and
$\{\Lambda^K_n(b)\}$.
In each case, the quantum
correction to the ground
state energy is the sum of the zero-point energies of the oscillators, 
\beq
\label{17}
\frac{1}{2h}\sum_n\sqrt{\Lambda_n},
\eeq
in units where $\hbar\equiv 1$ (recall that $h$ denotes the lattice spacing 
of the system).
 In the case of kinks, one should omit
from this sum the eigenvalue associated with the translation mode since,
the kink being very heavy, this mode is treated classically. Actually this 
makes no difference since the corresponding eigenvalue vanishes, so one
might as well sum over all modes, including the zero mode.

Of course, the series (\ref{17}) is divergent in both the vacuum and kink
sectors, and must be suitably regulated and renormalized. To this end, we
truncate the lattice symmetrically about the kink centre (so $-n_0\leq n\leq 
n_0$,
assuming $b\in[0,h)$) and consider the spectra 
$\{\Lambda_n(N):n=-1\ldots,N\}$
of the truncated $W$-matrices
of order $N=2n_0+1$. The renormalized ground state energy is then
\beq
\label{18}
{\cal E}(b)=\frac{E_P[\phi^{0,1}]}{\lambda^2}+
\lim_{N\ra\infty}\sum_{n=1}^{N}[\sqrt{|\Lambda_n^K(N,b)|}-
\sqrt{\Lambda_n^{vac}(N)}],
\eeq
which one expects to be finite, given the exponential spatial localization
of the kink (the large $|n|$ entries of the matrix $W^K$ are essentially 
identical to those of $W^{vac}$). The finite size of the lattice perturbs
the translation zero mode away from zero slightly, so one of the kink
eigenvalues $\Lambda_n(N,b)$ may be negative for some $(N,b)$ (although
it must vanish as $N\ra\infty$). This is why we have introduced an 
absolute value into equation (\ref{18}), so that ${\cal E}(b)$ is the limit
of a real valued sequence. In the limit $N\ra\infty$ lattice translation
symmetry is recovered, so ${\cal E}(b)$ should be periodic with period $h$.

\section{The $\cosh^{-\mu}$ kink family}
\label{family}

If we now choose $V=V_h$, for some kink profile $f$, we see from equations
(\ref{14}) and (\ref{4}) that the QPN potential depends on $f$ only through
$f'$, because
\beq
2+h^2V_h''(f(z))=\frac{f'(z+h)+f'(z-h)}{f'(z)}.
\eeq
For purposes of illustration we will consider the one-parameter family
\beq
\label{19}
f'(z)=\frac{1}{\cosh^\mu z}
\eeq
with $\mu>0$. Note that this includes the cases of the sine-Gordon ($f(z)=
2\tan^{-1}e^z$) and $\phi^4$ ($f(z)=\tanh z$) kink profiles: $\mu=1$ and
$\mu=2$ respectively. The corresponding transparent substrates in these two
cases are shown in figure 1.

\vbox{
\centerline{\epsfysize=2.2truein
\epsfbox[0 0 612 592]{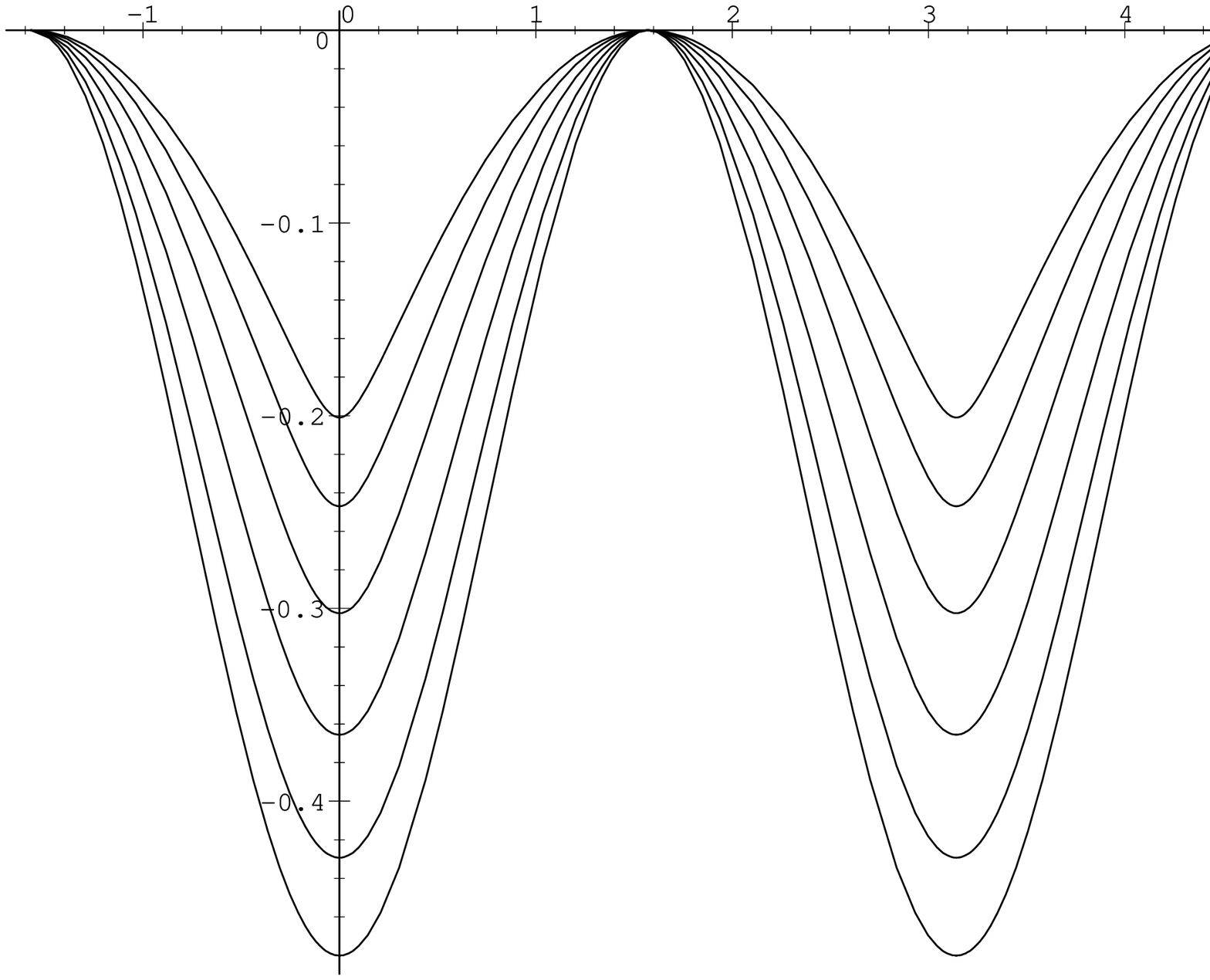}
\epsfysize=2.2truein
\epsfbox[0 0 612 592]{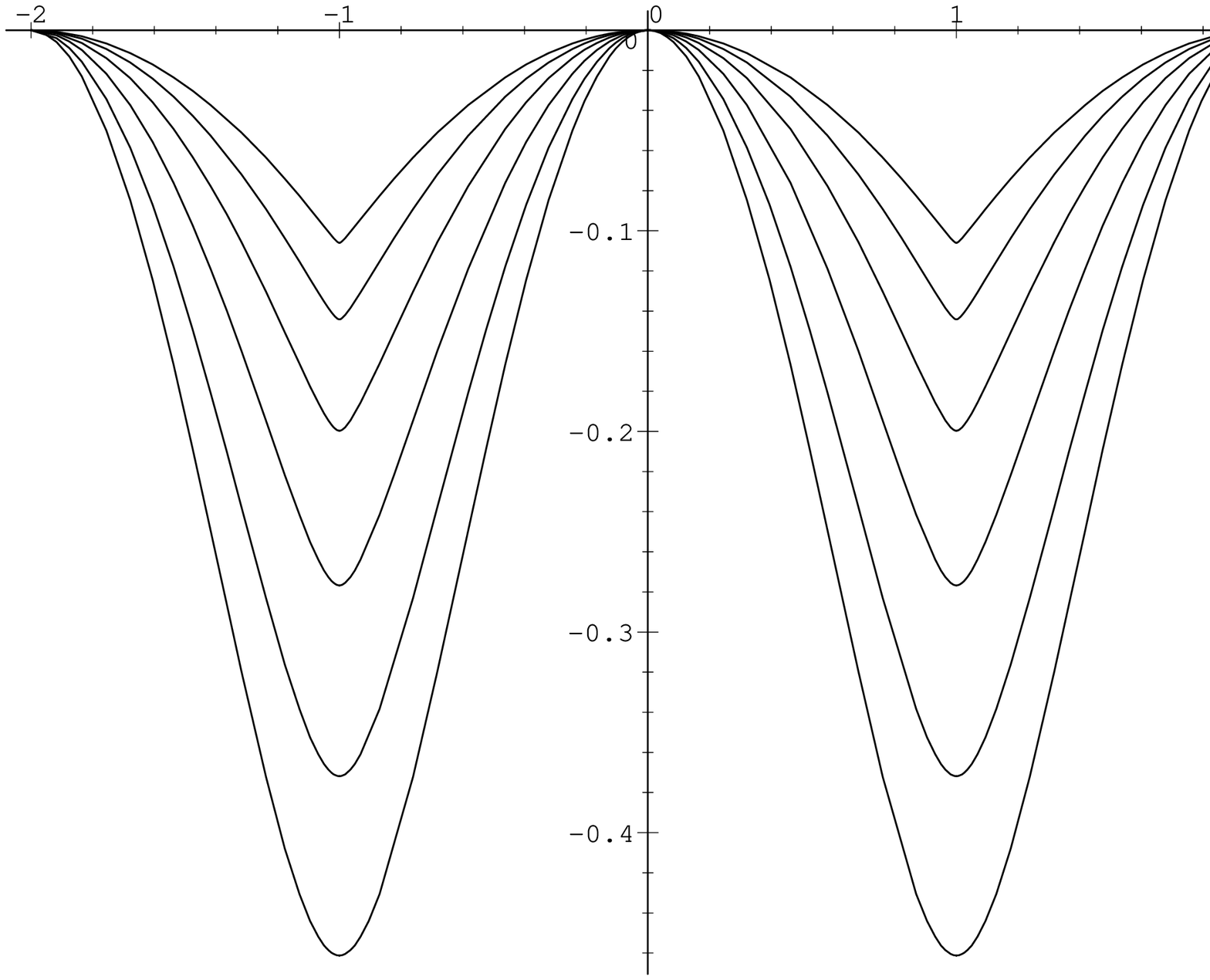}}
\noindent
{\it Figure 1: Transparent substrate potentials generated by the
sine-Gordon ($\mu=1$, left) and $\phi^4$ ($\mu=2$, right) kink profiles,
with lattice spacings from $h=0.5$ (bottom) to $h=3.0$ (top) 
in steps of $0.5$.}
}
\vspace{0.5cm}

The neighbouring vacua $a_\pm$ may be any real numbers separated by
\beq
a_+-a_-=\int_{-\infty}^\infty\frac{dx}{\cosh^\mu x}.
\eeq
Since $f(z)$ has order $e^{-\mu|z|}$ exponential decay, $V_h''(a_\pm)$ is
given by equation (\ref{6}), and the vacuum $W$-matrix takes the simple
form
\beq
\label{21}
W_{nm}^{vac}=2\cosh\mu h\, \delta_{nm}-(\delta_{n,m-1}+\delta_{n,m+1}).
\eeq
The spectrum of the system truncated to $N$ lattice sites is easily computed:
\beq
\Lambda_n^{vac}(N)=2(\cosh\mu h-1)+4\sin^2\left(\frac{n\pi}{2(N+1)}\right),
\qquad n=1,2,\ldots,N.
\eeq
In the limit $N\ra\infty$, the spectrum densely fills the interval
$[2(\cosh\mu h-1),2(\cosh\mu h+1)]$.

The eigenvalue problem for the kink W-matrix,
\beq
W^K_{nm}(b)=\frac{f'(nh+h-b)+f'(nh-h-b)}{f'(nh-b)}\, 
\delta_{nm}-(\delta_{n,m-1}+\delta_{n,m+1}),
\eeq
is intractable analytically, and will be solved numerically in section
\ref{numerics}. One can show, however,
that the quantum kink energy is (for all $b$) {\em lower} than the classical
kink energy, that is, the quantum energy correction is {\em negative}. To
see this, let $\Delta(b)$ be the real diagonal matrix
\beq
\Delta_{nm}(b)=h^2\delta_{nm}[V_h''(f(nh-b))-V_h''(a_\pm)],
\eeq
so that $W^K(b)=W^{vac}+\Delta(b)$. Let $\{\Lambda^K_n(b)\}$,
$\{\Lambda^{vac}_n\}$ and $\{\Gamma_n(b)\}$ be the eigenvalues of $W^K(b)$,
$W^{vac}$ and $\Delta(b)$, each spectrum arranged in nonincreasing order
($\Gamma_1\geq\Gamma_2\geq\Gamma_3\geq\ldots$). Then standard matrix 
perturbation theory \cite{minimax} asserts that
\beq
\Lambda_n^K(b)\leq\Lambda_n^{vac}+\Gamma_1(b)
\eeq
for all $n$, where $\Gamma_1$ is the greatest eigenvalue of $\Delta(b)$,
\beq
\Gamma_1(b)=\max_{n} h^2[V_h''(f(nh-b))-V_h''(a_\pm)].
\eeq
So if $V_h''(f(z))<V_h''(a_\pm)$ for all $z\in\R$ then $\Gamma_n<0$ and the
perturbation of $W^{vac}$ by $\Delta(b)$ must reduce each eigenvalue
$\Lambda_n^K(b)$ relative to its vacuum counterpart, with the result 
(always assuming kink stability) that
\beq
\sum_n(\sqrt{|\Lambda_n^K(b)|}-\sqrt{\Lambda_n^{vac}})<0.
\eeq

This condition on $V_h''(\phi)$ (maximum second derivative at the vacua)
is quite natural, and can easily be shown to hold for the whole 
$\cosh^{-\mu}$ family ($\mu>0$, $h>0$). Note that
\beq
h^2[V_h''(f(z))-V_h''(a_\pm)]=g(z)-\lim_{x\ra\infty}g(x),
\eeq
where
\beq
g(z)=\left[\frac{\cosh z}{\cosh(z+h)}\right]^\mu+
\left[\frac{\cosh z}{\cosh(z-h)}\right]^\mu.
\eeq
Now $g$ is differentiable and even, and
\beq
g'(z)=-\mu\sinh h\cosh^{\mu-1}z\left[\frac{1}{\cosh^{\mu+1}(z+h)}-
\frac{1}{\cosh^{\mu+1}(z-h)}\right],
\eeq
so the only critical point of $g$ is $z=0$, a local minimum, whence it
follows that $g(z)<\lim_{x\ra\infty}g(x)$ for all $z$.

\section{Numerical results}
\label{numerics}

Since the truncated kink $W$-matrix $W_N^K(b)$ is  real, symmetric and 
tridiagonal, its eigenvalue problem is particularly easy to solve 
numerically. In this section we present data generated by implementing the
QL decomposition algorithm for tridiagonal matrices with implicit eigenvalue 
shifts, outlined in \cite{recipes}.

The first thing to check is that, as expected, the spectrum of $W^K(b)$ is
positive semi-definite with nondegenerate eigenvalue zero. The least and
next-to-least eigenvalues of $W_N^K(b)$ for $N$ odd, $3\leq N\leq 90$ were
computed for a large sample of parameter values in the range
$1\leq\mu\leq 3$, $0.5\leq h\leq 10$, $0\leq b/h\leq 0.5$. The results were
similar at all points sampled: the least eignevalue converges to $0$, while 
the next-to-least converges to a positive number {\em below} the lower edge 
of the vacuum phonon band, that is, less than $2(\cosh\mu h-1)$. It is 
instructive to look at the build up of the spectrum of $W_N^K(b)$ as $N$
grows large, as depicted for two contrasting sets of parameter values in 
figure 2. This clearly shows the rapid convergence of the bottom eigenvalue
to $0$ (convergence being faster for larger $\mu h$, since the kink structure
is then more tightly spatially localized) and the next lowest eigenvalue to
a constant outside the phonon band. The rest of the eigenvalues accumulate,
apparently densely, within the phonon band (delimited by horizontal dashed
lines in figure 2). This is precisely the right behaviour to ensure both
kink stability and convergence of the quantum corrected kink energy in the
limit $N\ra\infty$. The size of truncated system needed to obtain practical
convergence depends on $\mu h$, but, within the parameter range cited above,
$N=51$ seems to suffice. This is the matrix order used to obtain the
remaining numerical data.

\vbox{
\centerline{\epsfysize=2.5truein
\epsfbox[59 201 551 620]{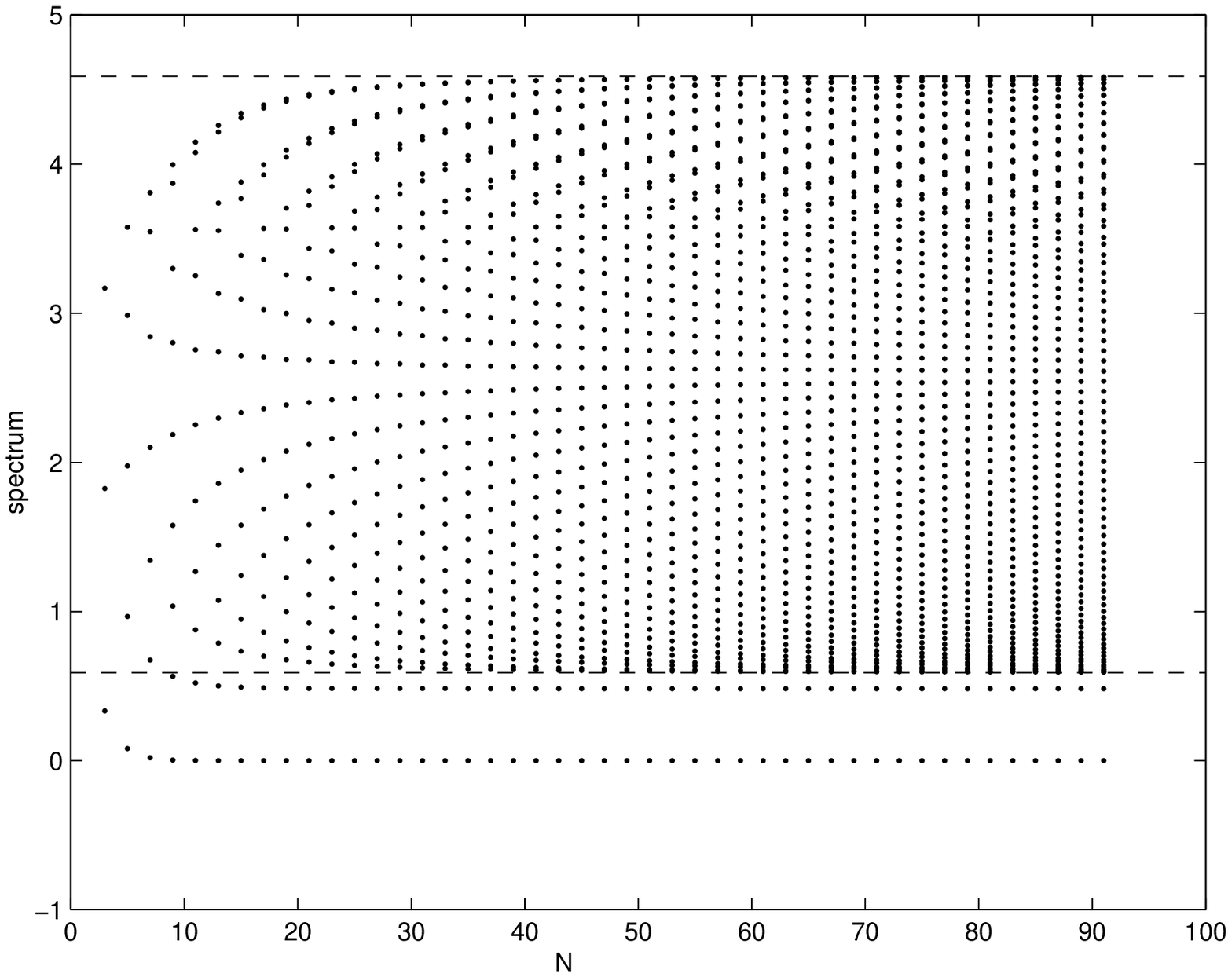}
\epsfysize=2.5truein
\epsfbox[59 201 551 620]{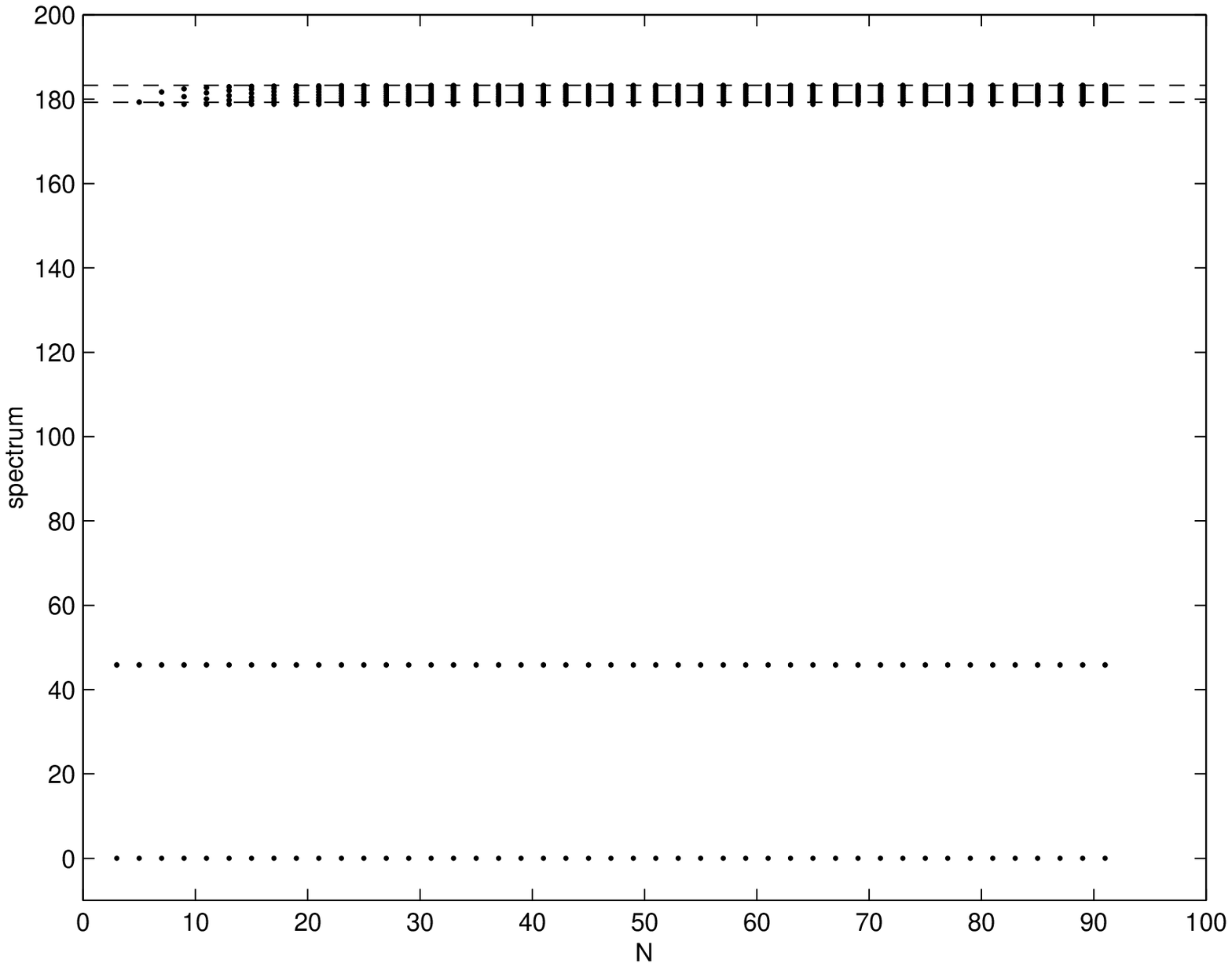}}
\noindent
{\it Figure 2: Build up of the spectrum of $W^K_N(b)$ as the matrix order
$N$ grows large, in the cases $\mu=1.5$, $h=0.5$, $b/h=0.2$ (left) and
$\mu=2.0$, $h=2.6$, $b/h=0$ (right). Note how the spectrum accumulates
within the vacuum phonon band, delimited by horizontal dashed lines.  }
}
\vspace{0.5cm}

The quantity ${\cal E}(b)$ as defined in (\ref{18}) is problematic to plot
since it contains contributions of different orders in $\lambda$. For this
reason it is convenient to consider $\et(b)
={\cal E}(b)-{\cal E}(0)$, which is 
of order $\lambda^0$, instead. In fact, $\et(b)$ is the quantity of most
direct physical significance anyway: it gives the change in kink energy as
$b$ varies, which is precisely what is meant by the quantum Peierls-Nabarro
potential.

Figure 3 shows $\et(b)$ ($0\leq b/h\leq 1$) for the sine-Gordon 
substrate potentials ($\mu=1$), 
at a variety of lattice spacings. The results are
qualitatively very similar to the usual classical PN potential: the kink
has greatest energy when located exactly on a lattice site ($b=0$) and least
when located exactly midway between lattice sites ($b=h/2$), the energy 
difference (the QPN barrier) growing monotonically with $h$. 

\vbox{
\centerline{\epsfysize=2.9truein
\epsfbox[50 201 546 620]{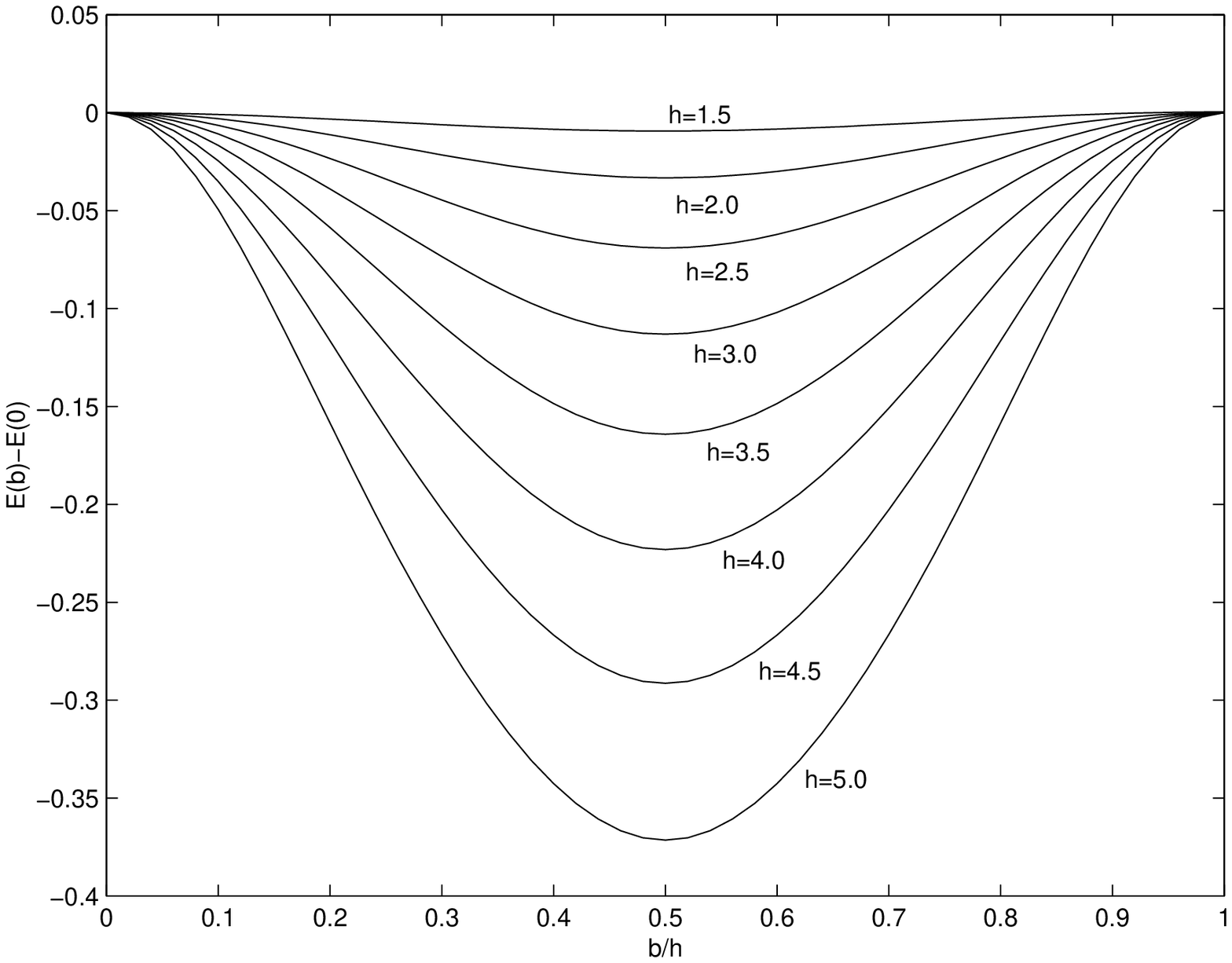}}
\noindent
{\it Figure 3: Position dependence of the quantum PN potential 
$\et(b)$ for $\mu=1$
and $h=1.5$ to $h=5$ in steps of $0.5$. }
}
\vspace{0.1cm}

\vbox{
\centerline{\epsfysize=2.9truein
\epsfbox[50 201 546 620]{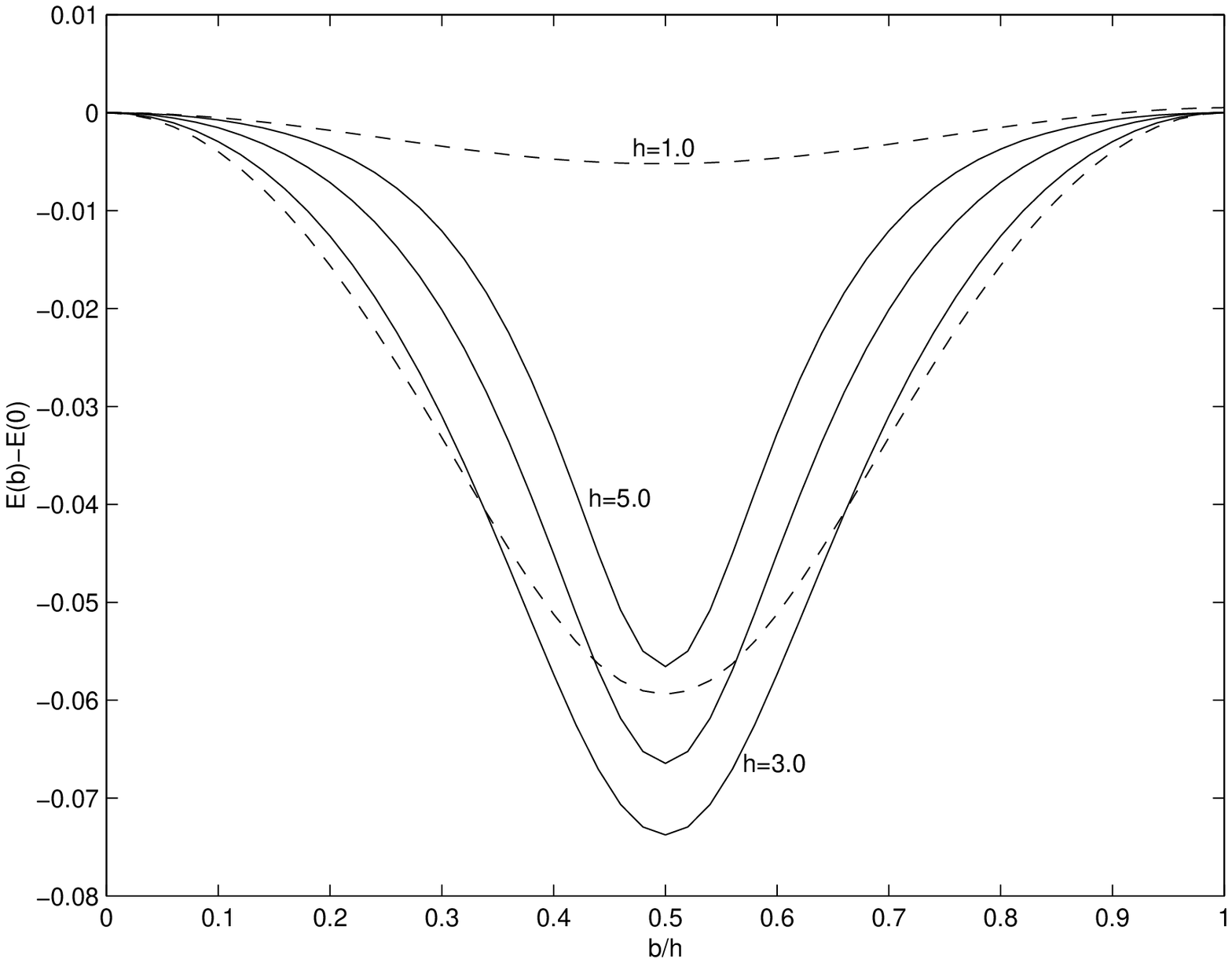}}
\noindent
{\it Figure 4: Position dependence of the quantum PN potential 
$\et(b)$ for $\mu=2$
and $h=2$ to $h=5$ in steps of $1$. The unlabeled curves are $h=2$ (dashed)
and $h=4$ (solid). }
}
\vspace{0.5cm}

Figure 4 shows similar plots for the $\phi^4$ substrate potentials ($\mu=2$).
Here, once again, kinks have greatest energy when $b=0$ and least when 
$b=h/2$, but the QPN barrier does not grow monotonically with $h$, nor
is the shape of $\et(b)$ so uniform as in the $\mu=1$ case.

Similar plots for $\mu=3$ reveal more complicated behaviour, as shown in 
figure 5.\ For small $h$, $\et(b)$ is maximum at $b=0$ and minimum at $b=h/2$
as for $\mu=1$, $\mu=2$. However, for $h$ above a critical value 
($\sim 1.46$) $b=0$ becomes a local minimum of $\et$ and two extra local
maxima appear. The global minimum of $\et$ remains at $b=h/2$, rather than
$b=0$, until $h$ exceeds about 1.52, after which $\et(h/2)$ exceeds 
$\et(0)$. As $h$ is increased further, the two local minima coallesce
at $b=h/2$ (at $h\sim 1.7$), so that the QPN potential starts to resemble an
inverted version of the $\mu=1$ case: now kinks have {\em minimum} energy
when located exactly on a lattice site and {\em maximum} energy when located
exactly midway between sites.

\vbox{
\centerline{\epsfysize=2.9truein
\epsfbox[50 201 546 620]{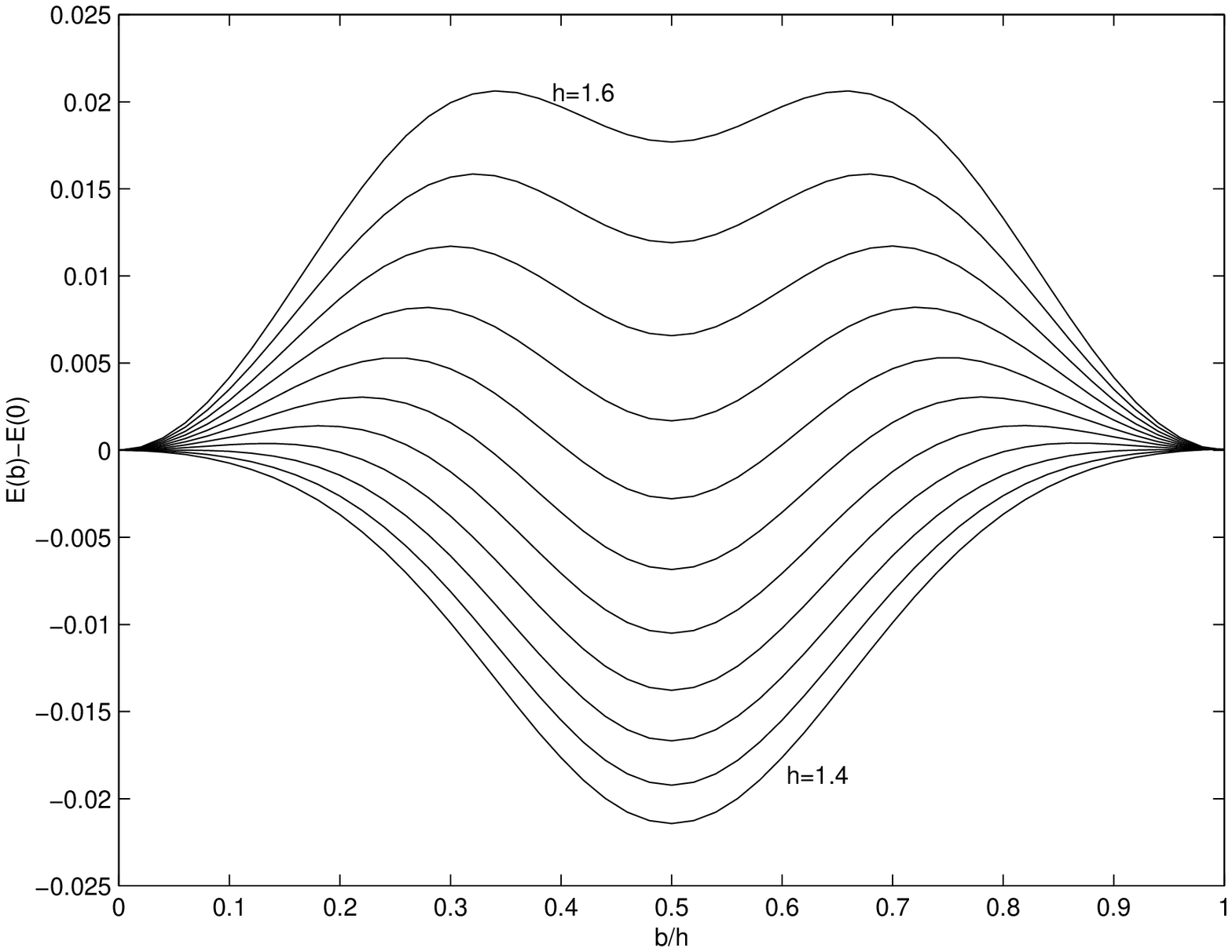}}
\noindent
{\it Figure 5: Position dependence of the quantum PN potential 
$\et(b)$ for $\mu=3$
and $h=1.4$ to $h=1.6$ in steps of $0.2$. }
}
\vspace{0.1cm}

\vbox{
\centerline{\epsfysize=2.9truein
\epsfbox[50 201 546 620]{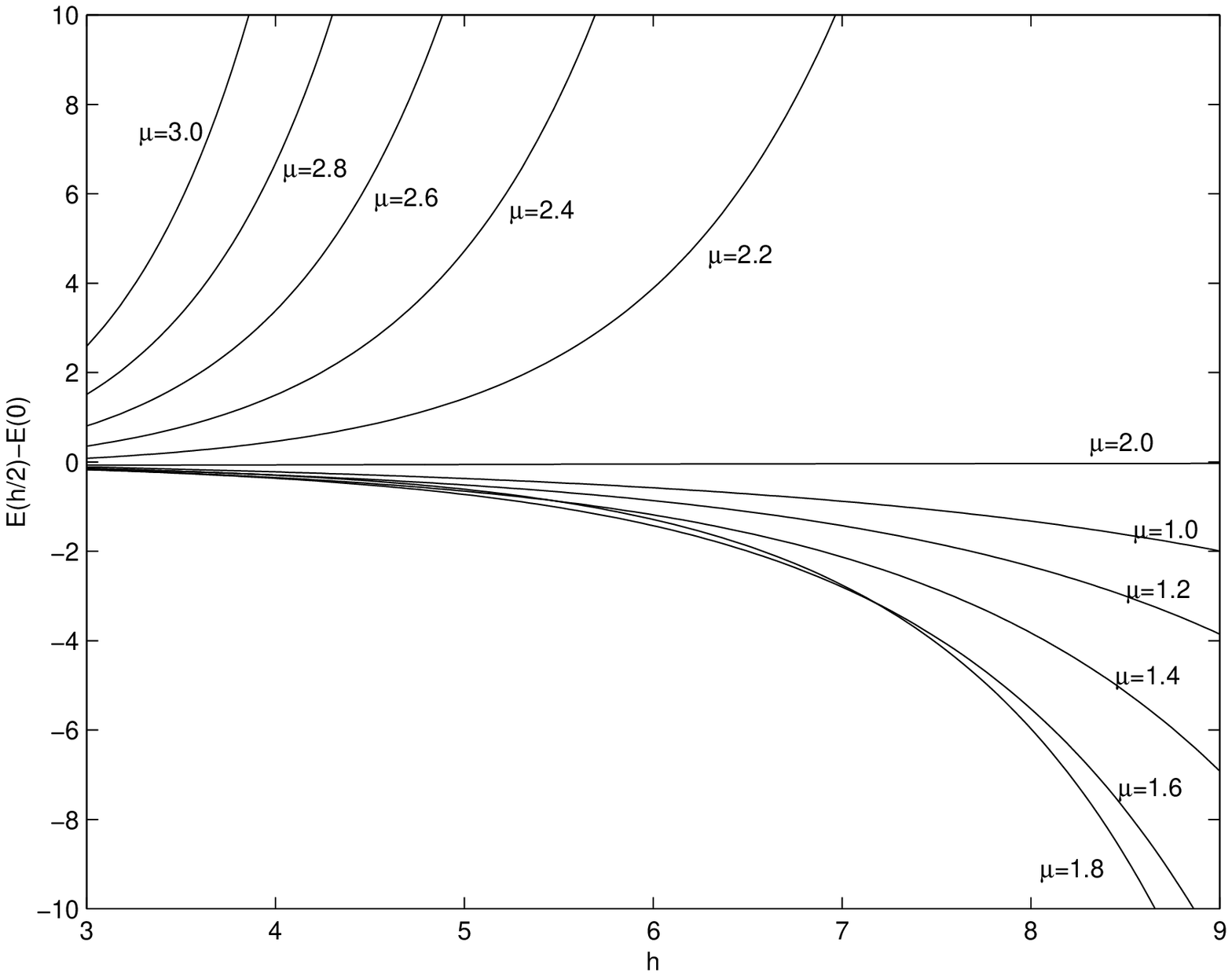}}
\noindent
{\it Figure 6: A rough measure of the depth of the QPN barrier:
$\et(h/2)$ as a function of $h$ for various $\mu\in[1,3]$. }
}
\vspace{0.5cm}

Periodicity and reflexion symmetry of ${\cal E}$ 
imply that $b=0$ and $b=h/2$ must
always be critical points. There may be others (as in the case $\mu=3.0$), 
but for the most part, plotting $\et(h/2)={\cal E}(h/2)-{\cal E}(0)$ against
$h$ gives a good account of how the QPN barrier varies with $h$ and $\mu$.
In particular, the sign of $\et(h/2)$ tells one whether the QPN barrier tends
to trap kinks between lattice sites ($\et(h/2)<0$) or on lattice sites
($\et(h/2)>0$).
Figure 6 shows plots of $\et(h/2)$ against $h$ for various $\mu\in[1,3]$.
Three regimes clearly emerge for large $h$: 
$1\leq\mu\leq 2$ ($\et(h/2)<0$ growing unbounded as $h\ra\infty$),
$\mu=2$ ($\et(h/2)<0$ remaining bounded as $h\ra\infty$) and
$2\leq\mu\leq 3$ ($\et(h/2)>0$ growing unbounded as $h\ra\infty$). This
trichotomy may be explained by consideration of the asymptotic forms of
$W^K(0)$ and $W^K(h/2)$ for large $h$:
\bea
\frac{W^K(0)}{e^{\mu h}}&\sim& {\rm diag}(\ldots,1,1,2^{-\mu},0,2^{-\mu},1,1,
\ldots) \\
\frac{W^K(h/2)}{e^{\mu h}}&\sim& {\rm diag}(\ldots,1,1,0,0,1,1,
\ldots).
\eea
This leads to the prediction
\bea
\frac{\et(h/2)}{e^{-\mu h/2}}&\sim&\frac{1}{2h}\left[(0-0)+(0-2^{-\mu/2})+
(1-2^{-\mu/2})+(1-1)+(1-1)+\ldots\right] \\
\label{29}
\Rightarrow
\et(h/2)&\sim&\frac{(1-2^{1-\mu/2})}{2h}e^{\mu h/2}.
\eea
Formula (\ref{29}) accounts well for the asymptotic behaviour seen in figure
6. Clearly the most interesting case from this point of view is the critical
value $\mu=2$. The dependence of $\et(h/2)$ on $h$ for $\mu=2$ is shown in
figure 7. One sees that, rather counterintuitively, the QPN barrier actually
vanishes in the extreme discrete limit, $h\ra\infty$. One should remember,
of course, that in varying $h$ one also varies the shape of the transparent
substrate potential $V_h$ (which would not otherwise remain transparent). In
fact, the limit $h\ra\infty$ is always
badly singular since the curvature of the
substrate at the vacua grows unbounded, by equation (\ref{6}).

\vbox{
\centerline{\epsfysize=2.9truein
\epsfbox[50 201 546 620]{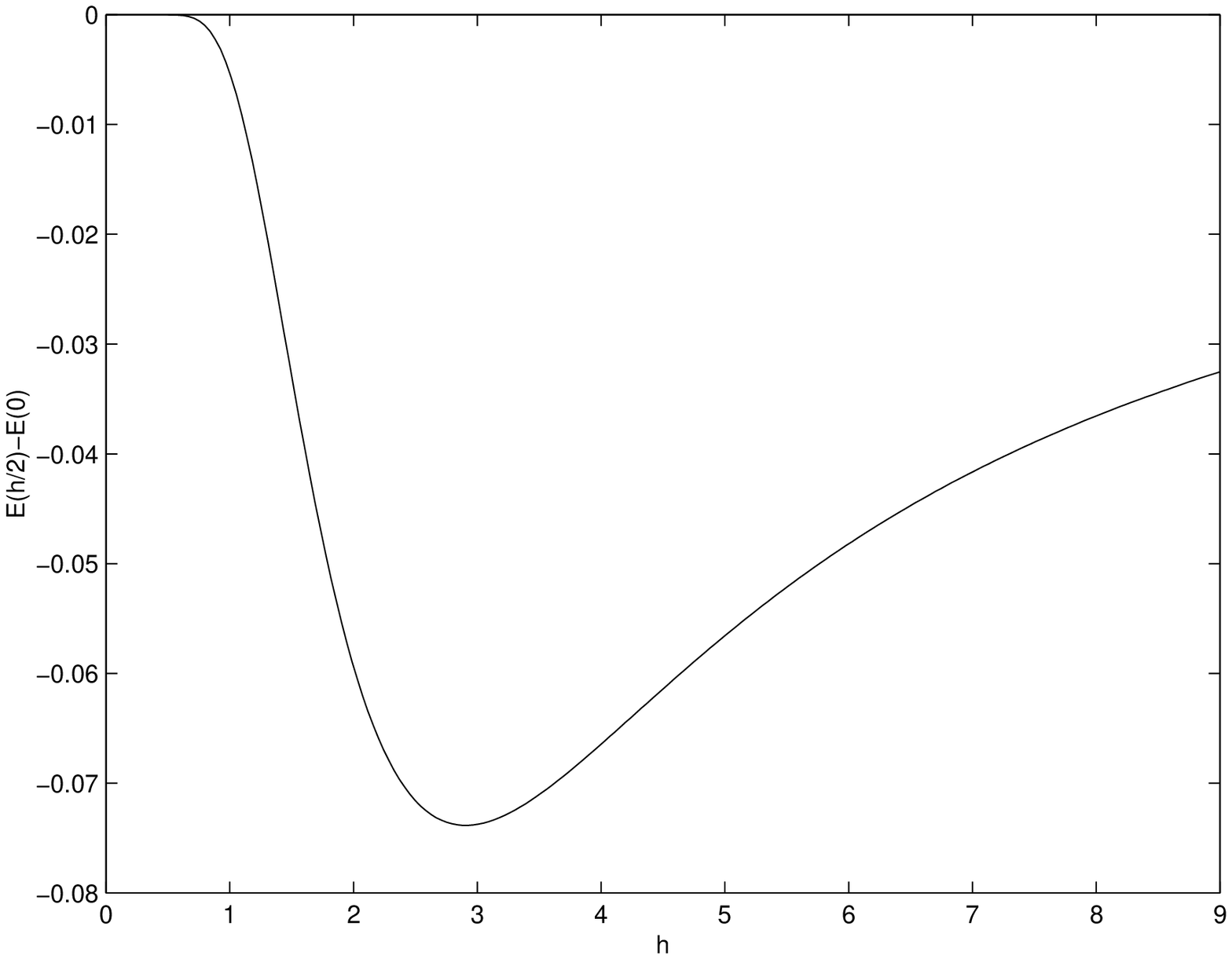}}
\noindent
{\it Figure 7: Dependence of the QPN barrier on the lattice spacing $h$ in
the critical case ($\mu=2$). }
}
\vspace{0.0cm}

\section{Concluding remarks}
\label{conc}

In this paper we have considered oscillator chains with no classical PN
barrier and shown that their kinks still experience a lattice-periodic
confining potential due to purely quantum mechanical effects, leading to a
new mechanism for kink pinning. The quantum PN potential was computed
numerically for a simple two-parameter family of substrates, revealing a rich
variety of behaviour.

It remains to be seen whether the QPN potential has any relevance to genuine
physics. Given the idealized one-dimensional nature of the model, this seems
unlikely (generalizing the Inverse Method to higher dimensions is very
problematic). Another cause for doubt is that the effect only exists for
certain special substrates. Just how special these ``transparent'' substrates
are is unknown. In order to have any physical relevance they would at least
have to be structurally stable: if $V_h[f]$ is the substrate transparent at 
$h$ generated by kink $f$, then given any sufficiently small perturbation 
$\delta V$ there should exist a kink $f_*$ close to $f$ and spacing $h_*$
close to $h$ such that $V_h[f]+\delta V=V_{h_*}[f_*]$. Thinking of the 
Inverse Method as a mapping $K\times\R_+\ra P$ (where $K$ is the space of
kink profiles and $P$ the space of potentials), the question is whether this
mapping is continuous with respect to some sensible choice of topologies on
$K$ and $P$. 

This is one of many open mathematical questions raised by the present work.
We have presented numerical evidence to support the assumption of classical
kink stability, but it should be possible to prove stability rigorously. 
Similarly, one should be able to prove convergence of the series defining
${\cal E}(b)$. In both cases one needs to understand the large $N$ behaviour
of the spectrum of $W^K_N(b)$. Standard minimax estimates are insufficient
for this purpose - more delicate analysis is required.

\section*{Acknowledgments}

This work was partially completed  at the 
Max-Planck-Institut f\"{u}r Mathematik in den Naturwissenschaften, Leipzig,
where the author
 was a guest scientist. He wishes to thank Prof.\ Eberhard Zeidler
for the generous hospitality of the institute. The author 
is an EPSRC Postdoctoral
Research Fellow in Mathematics.

\end{document}